\documentclass{aastex6}
\usepackage{amsmath}
\usepackage[normalem]{ulem}
\usepackage{appendix}
\def\numax{$\nu_{\rm max}$}
\def\nuac{$\nu_{\rm ac}$}
\def\dnu{$\Delta\nu$}
\def\teff{$T_{\rm eff}$}
\def\logg{$\log g$}
\def\mcs{${\mathcal S}$}
\def\be{\begin{equation}}
\def\ee{\end{equation}}

\begin{document}

\title{Changing the $\nu_{\mathrm{\MakeLowercase{max}}}$ Scaling Relation: The Need For a Mean Molecular Weight Term}

 
\author{Lucas S. Viani\altaffilmark{1}, Sarbani Basu\altaffilmark{1},  William J. Chaplin\altaffilmark{2,3}, Guy R. Davies\altaffilmark{2,3}, and Yvonne Elsworth\altaffilmark{2,3}}
 
\altaffiltext{1}{Department of Astronomy, Yale University, New Haven, CT, 06520, USA}

\altaffiltext{2}{School of Physics and Astronomy, University of Birmingham, Edgbaston, Birmingham, B15 2TT, UK}

\altaffiltext{3}{Stellar Astrophysics Centre (SAC), Department of Physics and Astronomy, Aarhus University, Ny Munkegade 120, DK-8000 Aarhus C, Denmark}

\email{lucas.viani@yale.edu}

\begin{abstract}
The scaling relations that relate the average asteroseismic parameters
$\Delta \nu$ and $\nu_{\mathrm{max}}$ to the global properties of stars are used quite 
extensively to determine stellar properties. While the $\Delta \nu$ scaling relation has been
examined carefully and the deviations from the relation have been well documented, the $\nu_{\mathrm{max}}$ scaling relation has not been examined as extensively. In this paper we examine the $\nu_{\mathrm{max}}$ scaling relation using a set of stellar models constructed
to have a wide range of mass, metallicity, and age. We find that as with $\Delta \nu$,
$\nu_{\mathrm{max}}$ does not follow the simple scaling relation. The most visible
deviation is because of a mean molecular weight term and  a $\Gamma_1$ term that are commonly ignored. The remaining deviation is more difficult to address. We find that the influence of the
scaling relation errors on asteroseismically derived values of $\log g$ are
well within uncertainties. The influence of the errors on mass and radius 
estimates is small for main sequence and subgiants, but can be quite large for red giants.
\end{abstract}


\section{Introduction}

The most easily determined asteroseismic parameters of a star are the large frequency separation, \dnu, and the frequency at which
oscillation power is maximum, \numax. These average asteroseismic parameters can
be determined even in poor signal-to-noise data, and 
as a result are commonly used in asteroseismic analyses. 

What makes \dnu\ and \numax\ so useful is that they are related to the
global properties of stars, their total mass, radius, and effective 
temperatures, through very simple relations, often known as scaling
relations.  The large frequency separation, \dnu, 
 is the average frequency spacing
between modes of adjacent radial order ($n$) of the same degree ($\ell$). 
The theory of stellar oscillations shows that  \citep[see, e.g.,][]{Tassoul1980,Ulrich1986, Christensen1993}
 \dnu\ scales approximately as the average density of a star, thus
\be
\Delta \nu \propto \sqrt{\bar{\rho}},
\label{eq:dnu1}
\ee
or in other words, we can approximate
\be
\frac{\Delta\nu}{\Delta\nu_\odot}\simeq \sqrt{\frac{M/M_\sun}{\left(R/R_\sun\right)^3}}.
\label{eq:dnu2}
\ee
The situation for \numax\ is a bit more complicated. As mentioned earlier,
\numax\ is the frequency at which oscillation power is maximum, and thus should
depend on how modes are excited and damped. Unlike the case of  \dnu, we do not as of yet
have a complete theory explaining the quantity. There are some studies in this regard
\citep[e.g., see][]{Belkacem2011,Belkacem2013}, but the issue has not been fully resolved. As explained in \cite{Belkacem2012pro} and \cite{Belkacem2011} the maximum in the power spectrum can be attributed to the depression or plateau of the damping rates. This is also discussed in \cite{Houdek1999}, \cite{Chaplin2008}, \cite{Belkacem2012}, and \cite{Appourchaux2012}. This depression of the damping rates can then be related to the thermal time-scale \citep{Balmforth1992,Belkacem2011,Belkacem2012pro} which in turn can be related to $\nu_{\mathrm{ac}}$, however there is some additional dependence on the Mach number  \citep{Belkacem2011,Belkacem2012pro}.

\numax\ carries diagnostic information on the excitation and damping of stellar modes, and hence must depend on the physical conditions in the near-surface layers where
the modes are excited. As assumed in \cite{Brown1991}, the frequency most relevant to these regions is the acoustic cut-off frequency, \nuac. The sharp rise in $\nu_{\rm ac}$ close to the surface of a star acts as an efficient boundary for the reflection of waves with $\nu < \nu_{\rm ac}$.  \citet{Brown1991} argued that \numax\ should be proportional to \nuac\ because  both frequencies are determined by conditions in the near surface layers. \citet{Kjeldsen1995} turned
this into a relation linking \numax\ to near-surface properties by noting that under the assumption of an isothermal atmosphere the acoustic-cutoff frequency can be approximated as 
\be
\nu_{\rm max} \propto \nu_{\rm ac} = \frac{c}{4\pi H},
\label{eq:nuac1}
\ee
where $c$ is the speed of sound and $H$ the density scale height (which under
this approximation is also the pressure scale height).
This can be further simplified assuming ideal gas as
\be
\nu_{\rm max} \propto \nu_{\rm ac} \propto g T^{-1/2}_{\rm eff},
\label{eq:nuac}
\ee
where $g$ is the acceleration due to
gravity and \teff\ the effective temperature. This leads to the \numax\ scaling
relation 
\be
\frac{\nu_{\mathrm{max}}}{\nu_{\mathrm{max},\sun}}=\left(\frac{M}{M_{\sun}}\right)
\left(\frac{R}{R_{\sun}}\right)^{-2} \left(\frac{T_{\rm eff}}{T_{{\rm eff},\sun}}\right)^{-1/2}.
\label{eq:numax}
\ee
While the \numax\ scaling relation and the relation between \numax\ and $\nu_\mathrm{ac}$ have not been tested extensively, limited observational studies as well as investigations using stellar models have been performed, suggesting that the approximations are reasonable \citep[e.g.][]{Bedding2003, Chaplin2008, Chaplin2011, Stello2008, Stello2009b, Miglio2012solo, Bedding2014, Jimenez2015, Coelho2015}. 

Equations~\ref{eq:dnu2}\ and \ref{eq:numax}\ have been used extensively, directly or
indirectly, to determine
the surface gravity, mass, radius, and luminosity of stars 
\citep[e.g.][etc.]{Stello2008, Bruntt2010, Kallinger2010b, Mosser2010, Basu2011, Chaplin2011, SilvaAguirre2011, Hekker2011direct, Chaplin2014, Pinsonneault2014}. Estimates of stellar properties may be determined from Eqs. \ref{eq:dnu2}\ and \ref{eq:numax}\
by treating them as two equations with two unknowns (assuming \teff\ is known 
independently) which leads to 
\begin{equation}
\frac{R}{R_\sun} = \left(\frac{\nu_{m\rm ax}}{\nu_{{\rm max},\sun}}\right) 
\left(\frac{\Delta \nu}{\Delta \nu_{\sun}}\right)^{-2} 
\left(\frac{T_{\mathrm{eff}}}{T_{\mathrm{eff,\sun}}}\right)^{1/2}
\label{eq:radius}
\end{equation}
and
\begin{equation}
\frac{M}{M_\sun} = \left(\frac{\nu_{\rm max}}{\nu_{{\rm max},\sun}}\right)^{3} 
\left(\frac{\Delta \nu}{\Delta \nu_{\sun}}\right)^{-4} 
\left(\frac{T_{\rm eff}}{T_{{\rm eff},\sun}}\right)^{3/2}.
\label{eq:mass}
\end{equation}
Determining the mass and radius of a star in this manner is known as the
``direct'' method. Comparisons with radius determinations made by other
techniques shows that the asteroseismic radii determined from the scaling relations
using the direct method hold to about 5\% for subgiants, dwarfs, and giants \citep{Bruntt2010, Huber2012, SilvaAguirre2012, Miglio2012, Miglio2012solo, White2013}. Examining red giants within eclipsing binaries, \cite{Gaulme2016} found radii determined using the direct method to be about 5\% too large. These deviations motivate us to investigate the \numax\ scaling relation since the $\Delta \nu$ scaling relation deviations have already been examined. 

Masses determined using Eq.~\ref{eq:mass} have uncertainties on the order of 10--15\% \citep[][and references therein]{Miglio2012solo, Chaplin2013,Chaplin2014}, however it is more difficult to test the masses given that there are few binaries with asteroseismic data. \cite{Brogaard2012} used eclipsing binaries in NGC 6791 and found the mass of red giant stars to be lower than the mass derived from studies which used the standard scaling relations \citep{Basu2011,Miglio2012,Wu2014}. \cite{Gaulme2016} found that for red giant stars within eclipsing binaries the direct method overestimated mass by around 15\%. \cite{Epstein2014} examined 9 metal-poor ($\mathrm{[M/H]}<-1$), $\alpha$-rich red giant stars and found their masses calculated from the scaling relation to be higher than expected. Also examining $\alpha$-enriched red giants, \cite{Martig2015} determined masses using the scaling relations and stellar models. The lower mass limit for each red giant was then converted into a maximum age. Unexpectedly, \cite{Martig2015} found a group of stars in the sample that were both young and $\alpha$-rich. Additionally, \cite{Sandquist2013}, studying NGC 6819, found that red giant masses in the cluster from asteroseismology are as much as 8\% too large while \cite{Frandsen2013} found indications that the detached eclipsing binary KIC 8410637 red giant star's mass was less than asteroseismology indicated. These results further indicate that there are uncertainties with the direct method and that the scaling relations in Eqs.~\ref{eq:dnu2} and \ref{eq:numax} need to be carefully understood.

Due to the wide use of the scaling relations in asteroseismology, the accuracy
of the $\mathrm{\nu_{max}}$ and $\Delta \nu$ scaling relations are crucial to
obtain a better understanding of stellar properties.  The scaling relations are
a result of approximations, and are therefore not expected to be completely accurate. The deviation of the \dnu\ scaling relation has been studied quite
extensively \citep[e.g.][]{White2011, Miglio2013_recomended, Mosser2013, Sharma2016, Guggenberger2016, Rodrigues2017}. It has been shown that the relation $\Delta\nu\propto \sqrt{\bar{\rho}}$ holds only to
a level of a few percent, and that $\Delta\nu/\sqrt{\bar{\rho}}$ instead of
being equal to unity is  a function of \teff\ and metallicity. At low \logg,
there also seems to be a dependence of $\Delta\nu/\sqrt{\bar{\rho}}$ on mass.
It is easy to get rid of this error in stellar models, all that one needs to do is
calculate \dnu\ for the models using the oscillation frequencies, rather than
Eq.~\ref{eq:dnu2}.  There are two approaches that can be used to account for
\dnu\ errors in the direct method. One is to ``correct'' the observed \dnu\
using a correction determined from models \citep{Sharma2016} and the other is to
use a temperature and metallicity dependent reference \dnu\ instead of the
solar value of \dnu\ in Eq.~\ref{eq:dnu2}. \citet{Yildiz2016}\ claimed that the
deviation of  $\Delta\nu/\sqrt{\bar{\rho}}$ from unity could be a result of
changes in the adiabatic index $\Gamma_1$, as this would affect the sound travel time. They found a linear relationship between $\Delta\nu/\sqrt{\bar{\rho}}$ and $\Gamma_1$ which they used to tune the scaling relation.

Unlike the \dnu\ scaling relation, the \numax\ scaling relation
has not been tested as extensively. Additionally, the tests have been indirect. \citet{Coelho2015} tested the temperature dependence of the \numax\ scaling relation
for dwarfs and subgiants and determined the classical $gT_{\rm eff}^{-1/2}$ scaling held to $\simeq$1.5\% over the 1560 K range in \teff\ that was tested. \citet{Yildiz2016}\ examined the $\Gamma_1$ dependence on \numax\ and found that the inclusion of a $\Gamma_1$ term alone, from the derivation of $\nu_{\rm ac}$, did not improve mass and radius estimates calculated using the scaling relations and in fact made mass and radius estimates worse than the traditional scaling relations (this is examined further in Sec.~\ref{sec:results}). \cite{Yildiz2016} found that additional tuning of the scaling relations (as a function of $\Gamma_1$) was needed. Other tests of the \numax\ scaling relation depend on comparing the radius and mass results obtained by using Eqs.~\ref{eq:dnu2}\ and \ref{eq:numax} with those obtained from either detailed modeling of stars \citep{Stello2009b, Silva2015} or of independently determined masses and radii \citep[e.g.,][]{Bedding2003, Bruntt2010, Miglio2012solo, Bedding2014}.

In addition to the inaccuracies in the scaling relations, there is another problem with using Equations~\ref{eq:radius}\ and \ref{eq:mass} in stellar radius and mass determination. The basic equations (Eqs.~\ref{eq:dnu2} and \ref{eq:numax}) that link \dnu\ and \numax\ to the mass, radius, and temperature of a star assume that all values of \teff\ are possible for a star of a given mass and radius.  However, the equations of stellar structure and evolution tell us otherwise --- we know that for a given mass and radius, only a narrow range of temperatures are allowed.
Additionally, we know that the mass-radius-temperature relationship depends on
the metallicity of a star; the scaling relations do not account for that. Thus,
an alternative to using Eqs.~\ref{eq:radius}\ and \ref{eq:mass} is to perform a
search for the observed  \dnu, \numax, \teff, and metallicity in a fine grid of
stellar models and to use the properties of the models to determine the
properties of the star. This is usually referred to as ``Grid Based Modeling''
(GBM) though it is more correctly a grid-based search and has been used
extensively to determine stellar parameters \citep[e.g.,][]{Chaplin2014,
Pinsonneault2014, Rodrigues2014}.  There are many different schemes that have been used for
GBM \citep[e.g.,][]{Stello2009, Basu2010, Quirion2010, Kallinger2010, Gai2011, Miglio2013, Hekker2013, Creevey2013, Serenelli2013}. While grid-based methods give more accurate results, they can give rise to model dependencies. Whether one uses the direct method to estimate masses, radii, and \logg, or used GBM, the results can only be as correct as the scaling relations.


One of the most important applications of the asteroseismic scaling relations
has been in estimating the surface gravity of stars. Spectroscopic 
surface-gravity measurements are notoriously difficult and inaccurate
and affect metallicity estimates.
It is becoming quite usual to use asteroseismic \logg\ values as priors
 before determining the metallicity from spectra
\citep[e.g.,][]{Bruntt2012, Brewer2015, Buchhave2015}.

In this paper we examine the \numax\ scaling relation in a similar manner as
to how the \dnu\ scaling relation has been tested. We use a set of stellar models
to do so. We should note from the outset that we are not testing the basic
assumption that $\nu_{\rm max}\propto \nu_{\rm ac}$, which is beyond the
scope of this paper,  but whether
$\nu_{\rm ac}$ (and hence \numax) follows the proportionality in
Eq.~\ref{eq:nuac}. We also examine the consequence of our results on asteroseismically
derived stellar properties, in particular, values of \logg, that are used so
widely.

The rest of the paper is organized as follows: we describe the models and
\numax\ calculations in Section~\ref{sec:models}, the results are presented
and discussed in Section~\ref{sec:results}. The consequences of the results are
discussed in Section~\ref{sec:conse} and we give some concluding remarks in
Section~\ref{sec:concl}.

\newpage
\section{Stellar Models and \numax\ calculations}
\label{sec:models}

\subsection{The models}
\label{subsec:models}

We use a grid of models to examine the \numax\ scaling relation. The models were constructed with the Yale  Rotating Evolutionary Code (YREC) \citep{Demarque2008} in its non-rotating configuration.
Models were created for seven different masses, $M=0.8$, 1.0, 1.2, 1.4, 1.6, 1.8, and 2.0 M$_\sun$ beginning at the zero-age main sequence through to the red giant branch. Models were stopped at the point where \numax\ of the models calculated using Eq.~\ref{eq:numax} was $3 \; \mu$Hz, where in Eq.~\ref{eq:numax} we adopted $\nu_{{\rm max},\odot}=3090 \; \mu$Hz. For each mass, models were constructed with  eight metallicities, $\mathrm{[Fe/H]}=$ $-$1.50, $-$1.00, $-$0.75, $-$0.50, $-$0.25, 0.0, 0.25, and 0.50. The grid does not include core helium burning stars. Two separate grids were constructed, one  with the Eddington T-$\tau$ relation in the atmosphere and one set with Model C of \citet{VALC1981} (henceforth referred to as the VAL-C atmosphere). In the latter case the atmosphere was assumed to be isothermal for $\tau \le 0.00014$ to avoid a temperature minimum. A temperature minimum is a feature that arises because of magnetic fields, which our models do not include. The models were constructed without the diffusion and gravitational settling of helium or heavy elements.

Both grids were constructed assuming
\citet{Grevesse1998} solar metallicities and
thus $\mathrm{[Fe/H]}=0$ is defined by $(Z/X)_\sun=0.023$. 
To determine $Y$ we first constructed two standard solar models (SSM), one with Eddington and
one with VAL-C atmosphere. The initial $Y$ and $Z$ needed to construct the SSM
was translated to the $Y$--$Z$ relation assuming that $Z=0$ when $Y$ has the
primordial value of 0.248. Since the construction of the SSMs yields solar-calibrated
mixing length parameters, those values were used to construct the models
of the grid ($\alpha_{\rm MLT}=1.70$ for Eddington models, 1.90 for VAL-C models).

The models were constructed using the OPAL equation of state \citep{Rogers2002}
and OPAL high-temperature opacities \citep{Iglesias1996} supplemented with 
low-temperature opacities of \citet{Ferguson2005}. Nuclear reaction rates of 
\citet{Adelberger1998} were used, except for that of the $^{14}N(p,\gamma)^{15}O$ reaction,
where the \citet{Formicola2004} rate was used.

\subsection{Calculating \numax}
\label{subsec:numax}

We calculate \numax\ for our models assuming that \numax\ is proportional to the acoustic cut-off frequency, which is the assumption that leads to the scaling relation in Eq.~\ref{eq:numax}. The acoustic cut-off frequency is the frequency above which waves are no longer trapped within the star. Waves of higher frequency form traveling waves and these high frequency ``pseudo modes'' are visible in power spectra. While below $\nu_{\mathrm{ac}}$ the modes are a sum of Lorentzians nearly equally spaced in frequency, above $\nu_{\mathrm{ac}}$ the pseudo mode shapes are more sinusoidal. These pseudo mode peaks are believed to be a result of interference between the waves that arrive at the observer having traveled different paths \citep[e.g.,][]{Kumar1991}. These high frequency waves can either travel directly towards the observer, leaving the star, or travel into the star before being reflected and leaving the star. Since the two waves travel different paths on their way to the observer, this results in constructive or destructive interference (depending on the path length difference and wavelength) creating peaks in the power spectrum. These pseudo modes can be used to observationally determine the acoustic cut-off frequencies of stars \citep[][etc.]{Garcia1998, Jimenez2015}.

\begin{figure*}
\epsscale{0.95}
\plotone{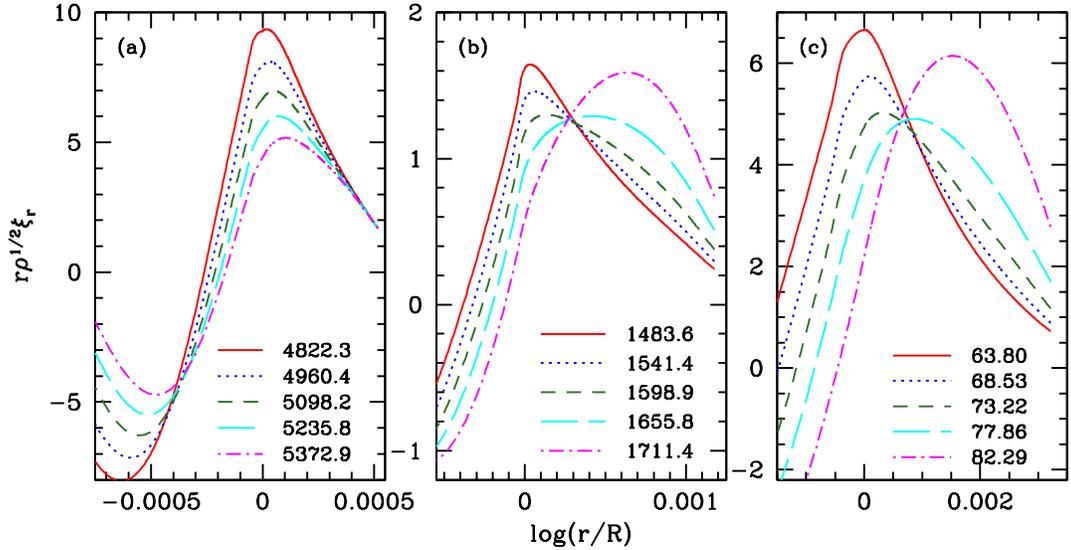}
\caption{The scaled eigenfunctions of modes with frequencies close to \nuac\ for
(a) a solar model, (b) model of a subgiant of mass 1.2 M$_\odot$ and $\mathrm{[Fe/H]}=-0.25$, and
(c) model of a red giant of mass 1.4 M$_\odot$ and $\mathrm{[Fe/H]}=0.25$. The legends indicate
the frequencies, in units of $\mu$Hz, that correspond to the eigenfunctions.
Note that the lower-frequency eigenfunctions in each case show a linear decay in $\log r$, the
higher frequency ones show a more oscillatory nature.
}
\label{fig:eig}
\end{figure*}
\begin{figure*}
\epsscale{0.95}
\plotone{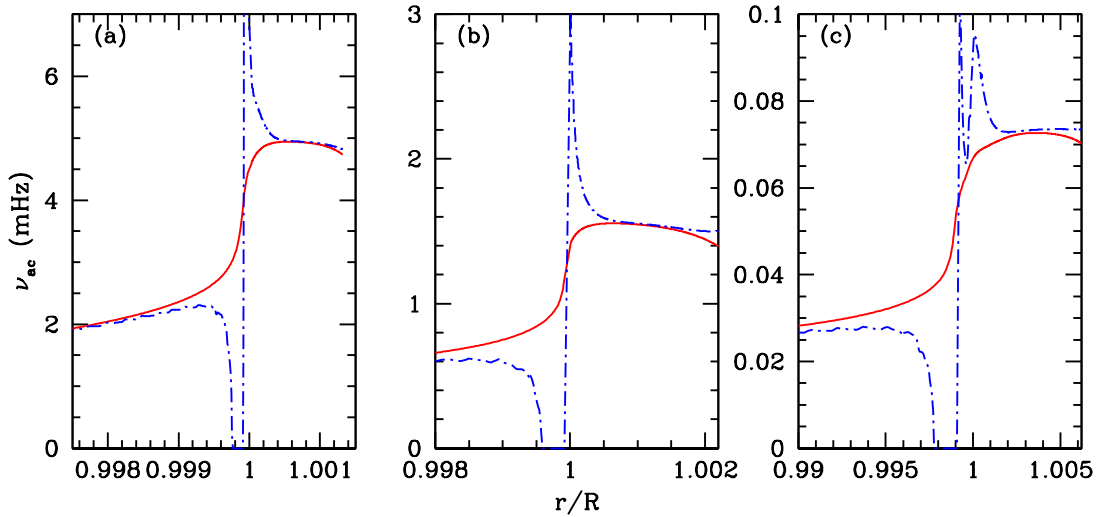}
\caption{The acoustic cut-off frequency for the three models shown in Fig.~\ref{fig:eig}\
calculated as per Eq.~\ref{eq:nuac1} (red solid line) and Eq.~\ref{eq:omac} (blue dot-dashed
line). The cut-off frequencies for the three models using Eq.~\ref{eq:nuac1} are
4.94 mHz, 1.55 mHz and 0.07 mHz respectively and quite consistent with the 
change in behavior of the eigenfunctions.
}
\label{fig:acous}
\end{figure*}

When it comes to the acoustic cut-off frequency of models, there are challenges. The acoustic cut-off divides modes into those that are trapped inside a star, and the pseudo modes that are not. In the former case, the displacement eigenfunctions 
decay in the atmosphere, in the latter they do not. However, there is no clear
boundary between the two, as is demonstrated in Fig.~\ref{fig:eig}. Thus 
when it comes to the acoustic cut-off frequency of models, 
one relies on an approximate theory \cite[see e.g.][]{Gough1993} that shows that the
acoustic cut-off is given by
\be
\nu_{\rm ac}^2=\frac{c^2}{16 \pi^2 H^2}\left(1-2\frac{dH}{dr}\right),
\label{eq:omac}
\ee
where $H$ is the density scale height.
In the case of an isothermal atmosphere, this reduces to the expression in
Eq.~\ref{eq:nuac1}. As is clear from both Eq.~\ref{eq:omac}\ and Eq.~\ref{eq:nuac1},
\nuac\ is a function of radius. The acoustic cut-off
of a model is assumed to be the maximum value of \nuac\ close to the
stellar surface. The acoustic cut-off frequencies defined by
Eq.~\ref{eq:omac} and Eq.~\ref{eq:nuac1}\ are reasonably similar (see Fig.~\ref{fig:acous}).
However, the frequency calculated using Eq.~\ref{eq:omac}\ has sharp changes close
to the top of the convection zone where large variations of the superadiabatic
gradient cause large changes in $\nu_{\rm ac}$, making it difficult to 
determine what the cut-off frequency should be. It is difficult to determine $\nu_{\mathrm{ac}}$ from the eigenfunctions, since the change from an exponential decay to an oscillatory nature is not sharp. However, they can guide us. Judging by the behavior of the eigenfunctions  shown in Fig.~\ref{fig:eig}\ and comparing the results with what we get as a maximum from Eq.~\ref{eq:nuac1}\ for the same models (Fig.~\ref{fig:acous}), using the isothermal approximation to calculate \numax\ should be adequate. In fact, this is what is usually done.

The \numax\ scaling relation is a proportionality and the Sun is used as the reference;
in other words for any given model, we can define the ratio
\be
R_{\rm sc}=\frac{\nu_{\rm max}}{\nu_{{\rm max},\odot}}=\left(\frac{M}{M_{\sun}}\right)
\left(\frac{R}{R_{\sun}}\right)^{-2} \left(\frac{T_{\rm eff}}{T_{{\rm eff},\sun}}\right)^{-1/2}.
\label{eq:ratscale}
\ee For each of our models, we can also define
\be
R_{\rm ac}=\frac{\nu_{\rm ac}}{\nu_{{\rm ac},{\rm SSM}}},
\label{eq:ratac}
\ee
where $\nu_{{\rm ac},{\rm SSM}}$ is the acoustic cut-off of a standard solar model
constructed with the same input physics (particular atmospheres) as the models.
The use of a solar model having the same physics to define the ratio allows us to
minimize effects related to improper modeling of the surface layers.
If the \numax\ scaling relation is perfect, the ratio
\be
{\mathcal S}=\frac{R_{\rm ac}}{R_{\rm sc}}
\label{eq:rat}
\ee
will be unity, if not, the scaling relation does not hold. We examine how
\mcs\ behaves in the next section. This \mcs\ parameter is the inverse of the $f_\nu$ parameter discussed in \cite{Yildiz2016}.

\section{Results}
\label{sec:results}

\begin{figure*}
\epsscale{0.95}
\plotone{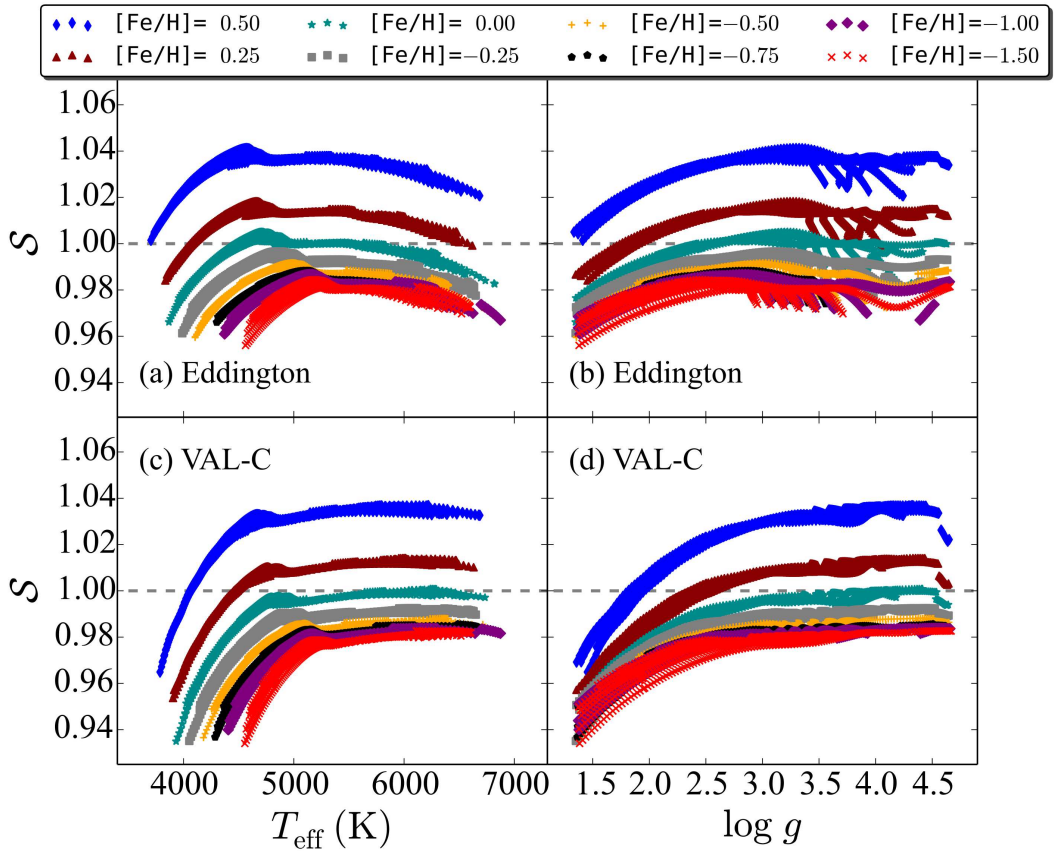}
\caption{The ratio \mcs\ (Eq.~\ref{eq:rat}) for the sets of  models with
Eddington atmospheres (top row) and VAL-C atmospheres (bottom row)
plotted as a function of \teff\ and \logg. The different colors and
symbols refer to different metallicities. The dashed gray line at ${\mathcal S}$=1 is provided for reference. Note the clear, systematic
offset that is a function of metallicity. 
}
\label{fig:s}
\end{figure*}

Figure~\ref{fig:s} shows the ratio \mcs\ plotted as a function of
\teff\ and \logg\ separately for the Eddington and VAL-C models.
Two features stand out immediately. First, that there is a metallicity
dependence which results in a systematic offset of \mcs\ for models
with non-solar metallicity. Secondly, that there is a deviation
at all metallicities at low \teff\ and low \logg, i.e., in evolved
models.

The origin of the metallicity dependence is easy to understand, and it is somewhat surprising that it has been neglected for so long, even in grid-based modelling of average asteroseismic data. To understand the effect we need to go back to the origin of the scaling relation.

Eq.~\ref{eq:nuac1} tells us that \nuac\ behaves as $c/H$. But
for an isothermal atmosphere $H=P/(\rho g)$. Since $c\propto\sqrt{P/\rho}$
then
\be
\nu_{\rm ac}\propto g\sqrt{\frac{\rho}{P}}.
\label{eq:gpr}
\ee
The assumption of an ideal gas law tell us that
\be
{\frac{P}{\rho}}={\mathcal R}\frac{T}{\mu},
\label{eq:ideal}
\ee
where ${\mathcal R}$ is the gas constant, and $\mu$ the mean molecular weight. Substitution of Eq.~\ref{eq:ideal} into Eq.~\ref{eq:gpr} gives
\be
\nu_{\rm ac}\propto g \sqrt{\frac{\mu}{T}}.
\label{eq:gprmu}
\ee
It should be noted that \citet{Jimenez2015} did include this term in their work.

\begin{figure*}
\epsscale{0.95}
\plotone{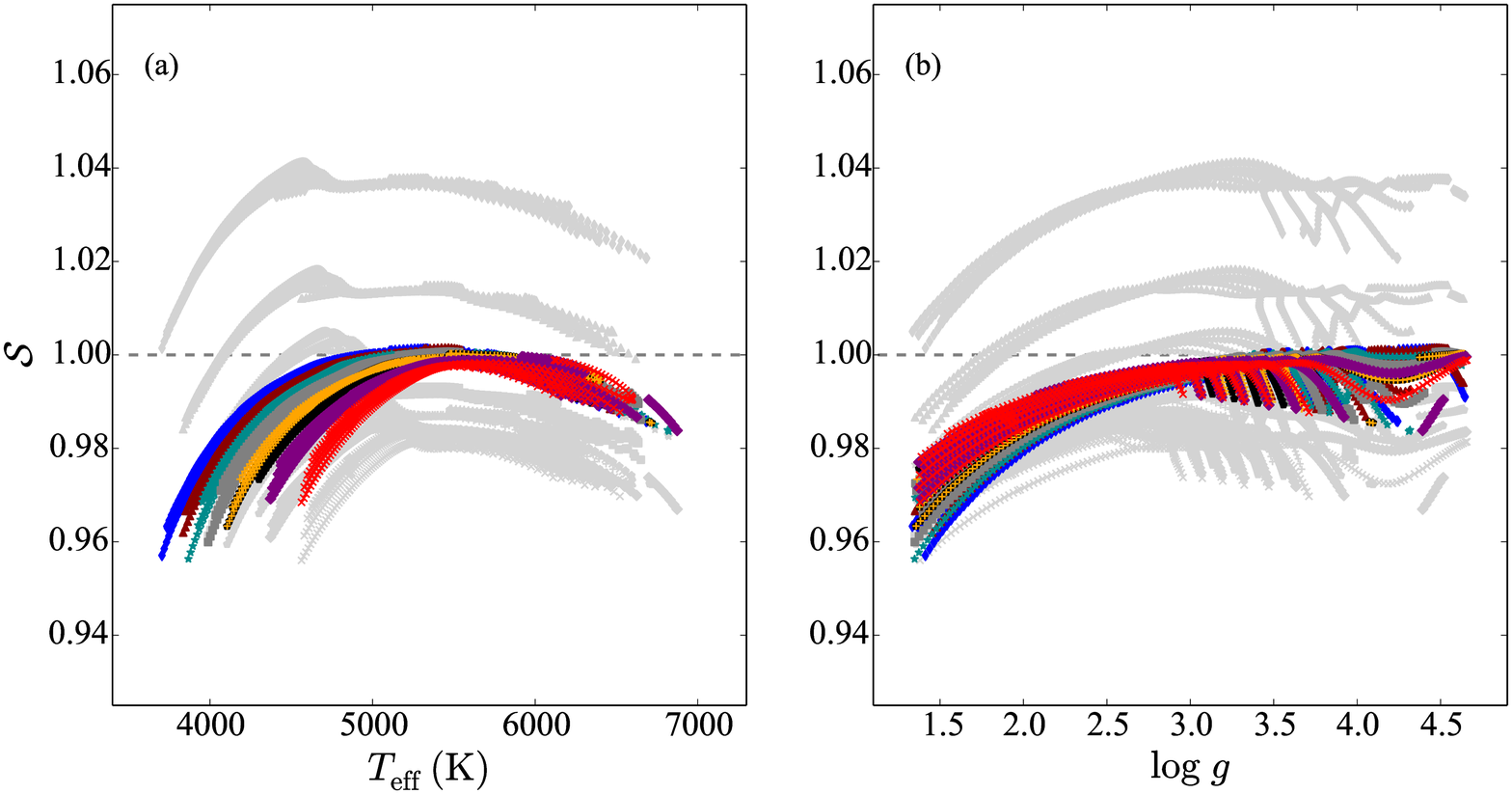}
\caption{The ratio \mcs\ calculated using Eq.~\ref{eq:numaxmu}\ to calculate \numax\ for the 
models with Eddington atmospheres plotted as a function
of \teff\ (a) and \logg\ (b). Symbols and colors correspond to metallicities as indicated in Fig.~\ref{fig:s}. The underlying light-gray points show the ratio \mcs\ calculated using the original scaling relation. Results for VAL-C atmospheres are similar and hence not shown.
Note that the systematic offset has disappeared, but there is still a remaining 
departure from the scaling relation.
}
\label{fig:mures}
\end{figure*}

Does the $\sqrt{\mu}$ term take care of the systematic difference seen for
the non-solar metallicity models? To test this we recalculated $R_{\rm sc}$
by modifying Eq.~\ref{eq:numax}\ to
\be
\frac{\nu_{\mathrm{max}}}{\nu_{\mathrm{max},\sun}}=\left(\frac{M}{M_{\sun}}\right)
\left(\frac{R}{R_{\sun}}\right)^{-2} \left(\frac{T_{\rm eff}}{T_{{\rm eff},\sun}}\right)^{-1/2}
\left(\frac{\mu}{\mu_\sun}\right)^{1/2}
\label{eq:numaxmu}
\ee
and calculated \mcs\ using the resultant modified $R_{\rm sc}$. The results are
shown in Fig.~\ref{fig:mures}. It can be seen clearly that the addition of the
$\sqrt{\mu}$ factor removes the difference between models with different
metallicities. One explanation for the usual omission of the $\mu$ term is that the abundances of $X$, $Y$, and $Z$ for an observed star, and therefore the value of $\mu$, can be difficult to determine. The application of the modified \numax\ scaling relation to observed stars will be discussed in Sec.~\ref{sec:concl}. While the importance of the $\mu$ term in the $\nu_{\mathrm{max}}$ scaling relation might seem to contradict what was found by \citet{Yildiz2016}, it should be noted that \citet{Yildiz2016} use models with a much smaller range in $\mu$ than the models presented in this work.

\begin{figure*}
\epsscale{0.95}
\plotone{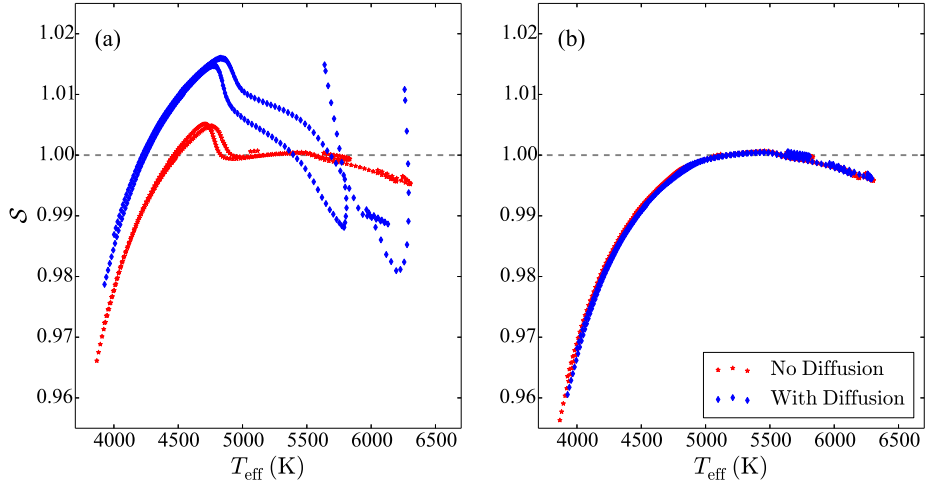}
\caption{A comparison of the ratio \mcs\ for models with and without diffusion
of helium and heavy elements. Panel (a) shows the results for the
original scaling relation while Panel (b) shows the results with the $\mu$-term included. For the sake of clarity only diffusion models of initial metallicity $\mathrm{[Fe/H]}=0.0$ are shown.
}
\label{fig:diffusion}
\end{figure*}

The main contribution to the difference in  mean molecular weight between
models with different metallicity is caused by differences in helium rather
than metals. This means that in models with diffusion
we should see a trend in the unmodified \mcs\ as a function of 
evolution that is different from that for models without diffusion. To test this we constructed Eddington models with diffusion for masses of $M=0.8$, 1.0, 1.2, and 1.4 $M_{\odot}$ and compared them to their corresponding non-diffusion models. The results, for diffusion models with initial metallicity of $\mathrm{[Fe/H]}=0.0$, are shown in Fig.~\ref{fig:diffusion}. As can be seen, the models with diffusion do indeed show a different trend, however, the trend disappears once the mean molecular weight is taken into account (see Fig.~\ref{fig:diffusion}(b)). Thus we conclude that if we are to use the \numax\ scaling relation, we need to explicitly use the $\mu$ dependence in the expression. The $\mu$ term could also be incorporated in the direct method provided that the model's $\mu$ value was known or could be calculated.

\begin{figure}
\epsscale{0.65}
\plotone{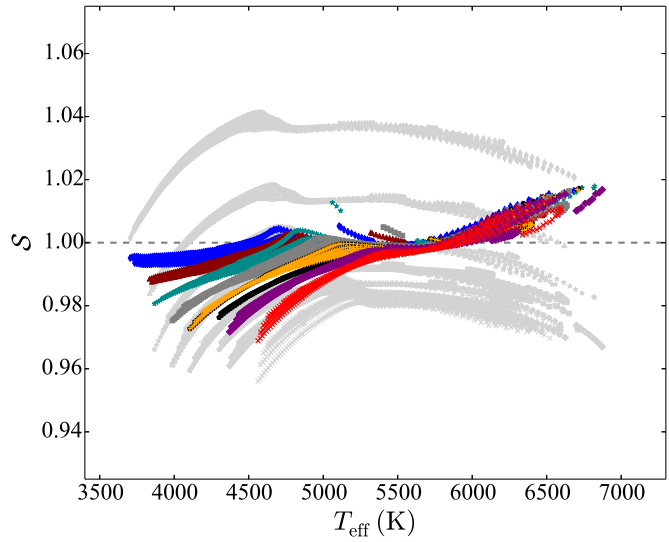}
\caption{The ratio  \mcs\ when the effects of both $\mu$ and $\Gamma_1$ are included. The underlying light-gray points show the ratio \mcs\ calculated using the original scaling relation. Symbols and colors correspond to metallicities as indicated in Fig.~\ref{fig:s}.}
\label{fig:gamma}
\end{figure}

Since $c^2=\Gamma_1 P/\rho$ we should also include a $\sqrt{\Gamma_1}$ in the scaling relation for \numax\ such that Eq.~\ref{eq:numaxmu} becomes 
\be
\frac{\nu_{\mathrm{max}}}{\nu_{\mathrm{max},\sun}}=\left(\frac{M}{M_{\sun}}\right)
\left(\frac{R}{R_{\sun}}\right)^{-2} \left(\frac{T_{\rm eff}}{T_{{\rm eff},\sun}}\right)^{-1/2}
\left(\frac{\mu}{\mu_\sun}\right)^{1/2} \left(\frac{\Gamma_1}{\Gamma_{1,\sun}}\right)^{1/2}.
\label{eq:numaxgamma}
\ee
Fig.~\ref{fig:gamma} examines the ratio \mcs\ when $R_{\mathrm{sc}}$ is calculated using Eq.~\ref{eq:numaxgamma}. As can be seen, the inclusion of the $\Gamma_1$ term also reduces the differences between the models of different metallicities and lessens the deviations from $ \mathcal{S} =1 $.

\begin{figure}
\epsscale{0.75}
\plotone{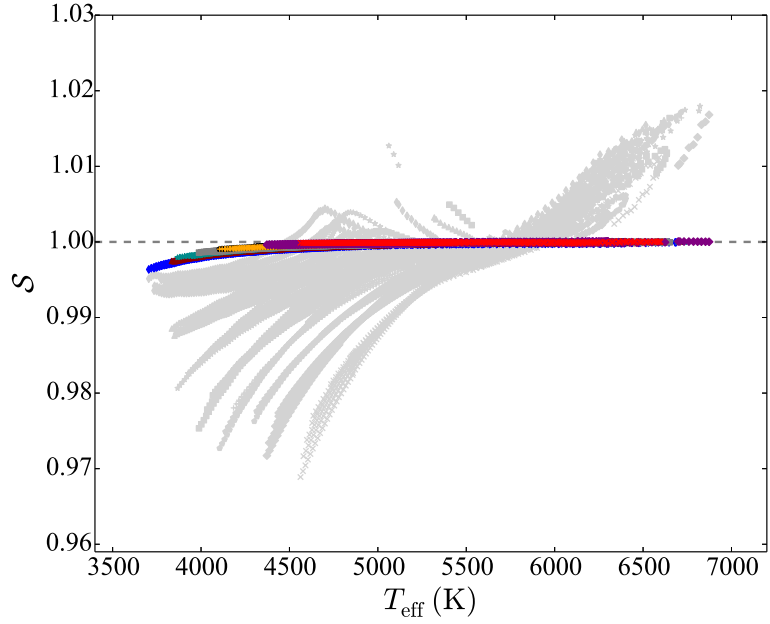}
\caption{The ratio \mcs\ calculated using Eq.~\ref{eq:numaxrad}\ to calculate \numax\ for the 
models with Eddington atmospheres. The underlying light-gray points show the ratio \mcs\ from Eq.~\ref{eq:numaxgamma}. Note that all systematic errors have been reduced. Symbols and colors correspond to metallicities as indicated in Fig.~\ref{fig:s}.
}
\label{fig:nuac}
\end{figure}

\begin{figure*}
\epsscale{0.95}
\plotone{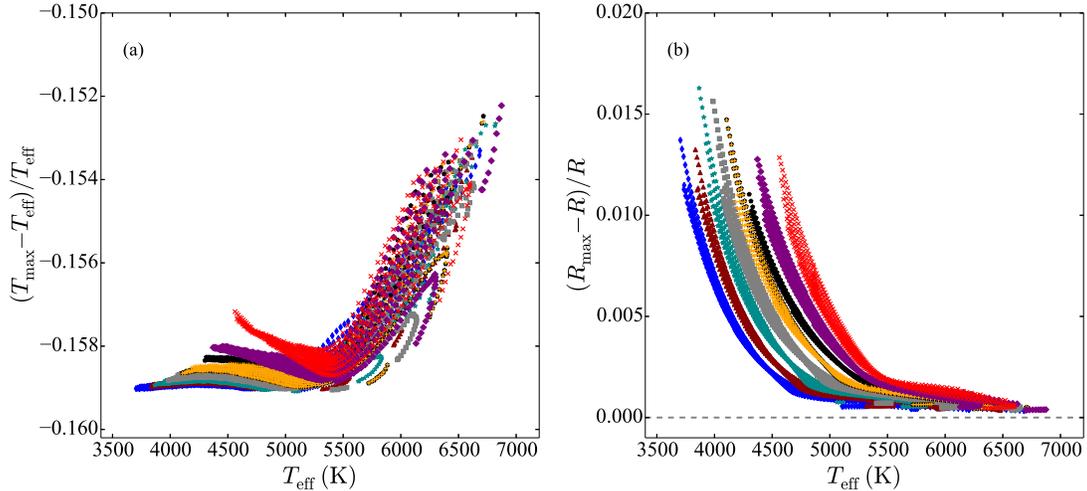}
\caption{The fractional differences between (a) $T_{\rm max}$ and $T_{\mathrm{eff}}$ and (b) $R_{\rm max}$ and $R$ for each model with an Eddington atmosphere. Symbols and colors correspond to metallicities as indicated in Fig.~\ref{fig:s}.
}
\label{fig:max_compare}
\end{figure*}

Even when the main deviation from the scaling relation is removed, there is a residual 
difference at low \teff. Most of this deviation can be explained by the fact that the maximum value of \nuac\ does not occur at $r=R$ but at a different radius. As seen in Fig.~\ref{fig:nuac}, this deviation is significantly lessened if Eq. \ref{eq:numaxgamma} is modified so that \numax\ is instead scaled as
\be
\frac{\nu_{\mathrm{max}}}{\nu_{\mathrm{max},\sun}}=\left(\frac{M_{\rm max}}{M_{{\rm max},\sun}}\right)
\left(\frac{R_{\rm max}}{R_{{\rm max},\sun}}\right)^{-2} \left(\frac{T_{\rm max}}{T_{{\rm max},\sun}}\right)^{-1/2}
\left(\frac{\mu_{\rm max}}{\mu_{\rm{max,\sun}}}\right)^{1/2} \left(\frac{\Gamma_{1,\rm max}}{\Gamma_{\rm{1,max,\sun}}}\right)^{1/2},
\label{eq:numaxrad}
\ee
where $R_{\rm max}$, $M_{\rm max}$, $T_{\rm max}$, $\mu_{\rm max}$, and $\Gamma_{\rm 1,max}$ are the radius, mass, temperature, mean molecular weight, and $\Gamma_1$
at the radius where $\nu_{\rm ac}$ is the maximum (note that for all models
$M_{\rm max}=M$ since the atmosphere is usually assumed to be massless), and
$R_{{\rm max},\odot}$, $M_{{\rm max},\odot}$, $T_{{\rm max},\odot}$, $\mu_{\rm max, \odot}$, and $\Gamma_{\rm 1, max, \odot}$ are the same quantities for a solar model with the same physics. As can be seen in Fig.~\ref{fig:nuac}, when taking into account that the maximum value of \nuac\ is not at r=R, the remaining deviation in the scaling relation is dramatically reduced. To compare the differences between the values of $T_{\rm max}$ and $T_{\mathrm{eff}}$ and $R_{\rm max}$ and $R$, refer to Fig.~\ref{fig:max_compare}.

\section{Consequences of the error on the \numax\ scaling relation}
\label{sec:conse}

Because the \numax\ scaling relation is used extensively along with the
\dnu\ scaling relation to estimate stellar properties, any deviations from
the scaling relation will add to systematic errors in the estimates. In this section, using the errors in the \numax\ scaling relation implied from our previous analysis, we examine the consequences on stellar \logg, radius, and mass estimates.

\begin{figure*}
\epsscale{0.95}
\plotone{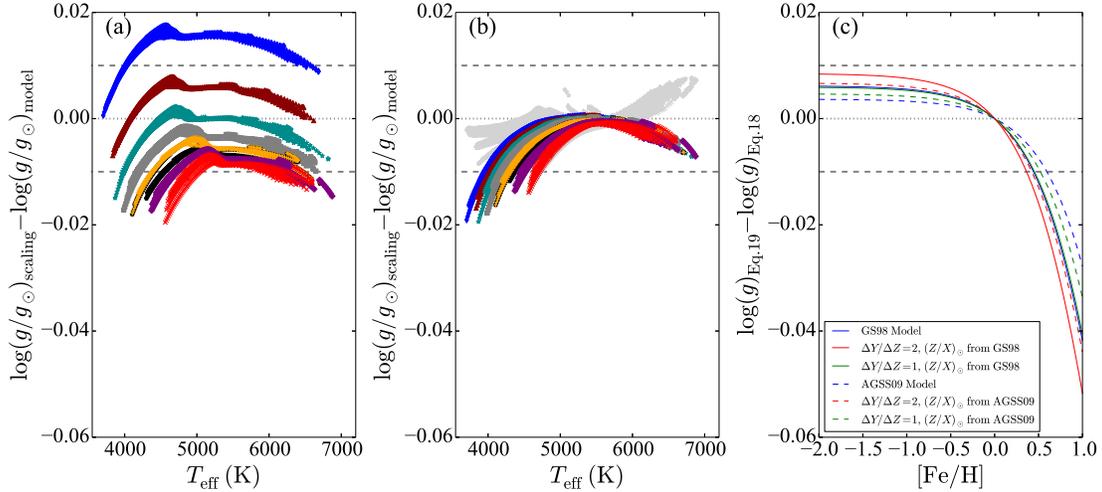}
\caption{(a) The error made in $\log g$ estimates when the original scaling relation
is used and (b) when the $\mu$ term is included (colored points) and when the $\mu$ and $\Gamma_1$ terms are included (background gray points). Symbols and colors correspond to metallicities as indicated in Fig.~\ref{fig:s}. (c) The difference between $\log g$ estimates if the $\mu$ term is included or not, plotted as a function of [Fe/H] for models with different values of  $\Delta Y/\Delta Z$. In each panel the gray dashed lines at $\pm$0.01 indicate typical uncertainties in asteroseismic $\log g$, with a dotted gray line at 0.0 for reference.
}
\label{fig:logdif}
\end{figure*}

As mentioned earlier, asteroseismic estimates of \logg\ are often
used as priors in spectroscopic analyses used to estimate atmospheric properties and parameters. Thus errors in asteroseismic estimates of \logg\ because of \numax\ errors is
a troubling matter. To test what systematic errors could result, 
we calculate \logg\ for the models from \numax\ using the usually
accepted relation for \numax, but with \numax\ of the models calculated
with the acoustic cut-off frequency, i.e.,
\be
\frac{g}{g_\odot}=\frac{\nu_{\rm ac}}{\nu_{{\rm ac}, {\rm  SSM}}}
\left(\frac{ T_{\rm eff}}{T_{{\rm eff},\odot}}\right)^{1/2},
\label{eq:logold}
\ee
and compare that to the actual \logg\ of the models. 
The results are shown in Fig.~\ref{fig:logdif}(a). As can be seen, there is
indeed a systematic error, but for the metallicity range of stars
for which asteroseismic \logg\ values have been measured, the
systematic error is well within the uncertainty range of data uncertainties (of the order $\pm 0.01$ dex). 
The systematic effects are somewhat larger in
the low temperature range that corresponds to red giants. This error can be
made much smaller if a $\mu$ term or a $\mu$ and $\Gamma_1$ term are included, i.e., if $g$ is calculated as
\be
\frac{g}{g_\odot}=\frac{\nu_{\rm ac}}{\nu_{{\rm ac}, {\rm  SSM}}}
\left(\frac{ T_{\rm eff}}{T_{{\rm eff},\odot}}\right)^{1/2}
\left(\frac{\mu }{\mu_\odot}\right)^{-1/2}
\label{eq:lognew}
\ee
or if also including the $\Gamma_1$ term,
\be 
\frac{g}{g_\odot}=\frac{\nu_{\rm ac}}{\nu_{{\rm ac}, {\rm  SSM}}}
\left(\frac{ T_{\rm eff}}{T_{{\rm eff},\odot}}\right)^{1/2}
\left(\frac{\mu }{\mu_\odot}\right)^{-1/2} \left(\frac{\Gamma_1 }{\Gamma_{1,\odot}}\right)^{-1/2}.
\label{eq:lognew_gamma}
\ee
The effects of calculating \logg\ using Eq.~\ref{eq:lognew} or ~\ref{eq:lognew_gamma} can be seen in Fig.~\ref{fig:logdif}(b). The addition of the $\mu$ term or the $\mu$ and $\Gamma_1$ terms lessens the deviations between models of different metallicities and brings the value of \logg\ from scaling more into agreement with the actual \logg\ values of the models.

In Fig.~\ref{fig:logdif}(c) we examine the difference in \logg\ estimates if the $\mu$ term is included or not as a function of [Fe/H]. We ignore the $\Gamma_1$ term here, as determining the value of $\Gamma_1$ for an arbitrary star with a given [Fe/H] is not as clear as determining $\mu$ for that star. So, in Fig.~\ref{fig:logdif}(c) we are examining the difference between \logg\ calculated with Eqs.~\ref{eq:logold} and \ref{eq:lognew}. Since the ratio $\mu/\mu_\odot$ depends on the $Y$--$Z$ relationship, we include different values of solar metallicity and different values of $\Delta Y/\Delta Z$ as a function of [Fe/H]. In Fig.~\ref{fig:logdif}(c) both the \citet{Grevesse1998} value of $(Z/X)_\sun=0.023$ (GS98) and the \citet{Asplund2009} value of $(Z/X)_\sun=0.018$ (AGSS09) are used. The added uncertainty because of the uncertainty in $\Delta Y/\Delta Z$ is small (less than $\pm 0.01$ dex) at low metallicity, but increases with an increase in metallicity. We are yet to gather asteroseismic data for stars with [Fe/H] larger than about 0.5, thus the errors for observed stars are expected to be quite low and smaller than typical \logg\ uncertainties. While this is a reassuring confirmation that the original scaling relation has produced trustworthy \logg\ estimates, Fig.~\ref{fig:logdif}(b) shows that \logg\ estimates are improved if Eqs.~\ref{eq:lognew} or \ref{eq:lognew_gamma} are used.
\begin{figure}
\epsscale{0.95}
\plotone{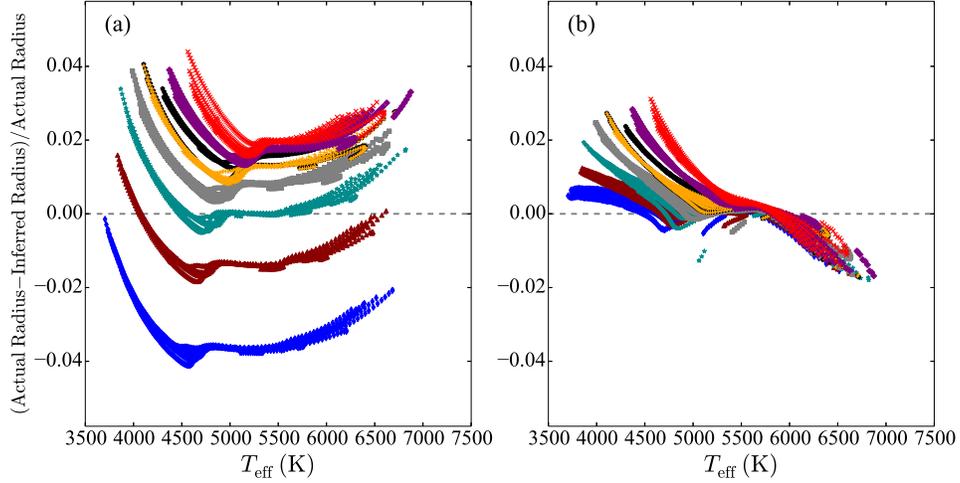}
\caption{(a) The error made in radius estimates when the original \numax\ scaling relation
is used. (b) The same when the \numax\ scaling relation is modified to include the $\mu$ and $\Gamma_1$ terms. Symbols and colors correspond to metallicities as indicated in Fig.~\ref{fig:s}. \dnu\ in both cases was calculated using the scaling relation to avoid introducing errors in the
radius estimates from the \dnu\ scaling errors.
}
\label{fig:raderr}
\end{figure}
\begin{figure}
\epsscale{0.95}
\plotone{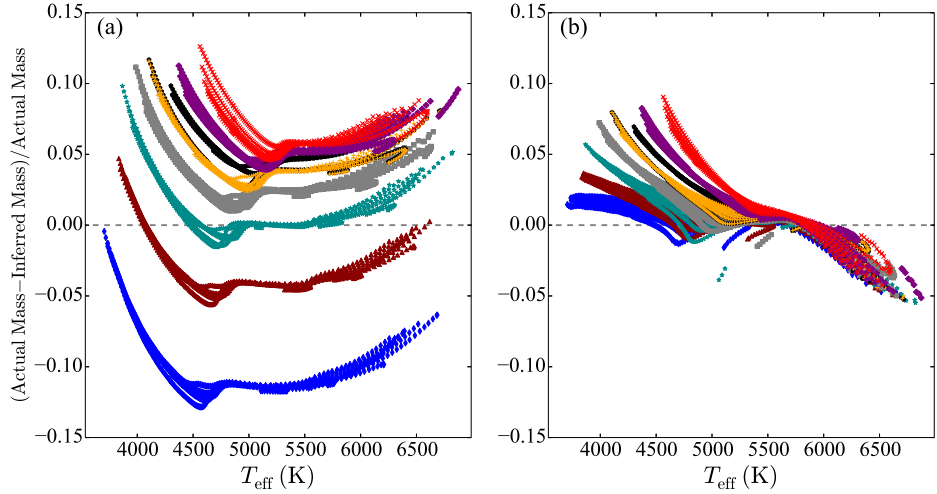}
\caption{Same as Fig.~\ref{fig:raderr} but shows errors in mass estimates.
}
\label{fig:masserr}
\end{figure}

What of the errors in radius and mass estimates that arise due to deviations in the \numax\ scaling relation? Using Eq.~\ref{eq:radius}\ and \ref{eq:mass} each model's radius and mass was determined, where $\nu_{\rm max} / \nu_{\rm max,\odot}$ was calculated using $\nu_{\rm ac} / \nu_{\rm ac,\odot}$. Here \dnu\ was calculated using the scaling relation in Eq.~\ref{eq:dnu2}, as opposed to calculating \dnu\ for each model using mode frequencies. This was done in order to avoid introducing errors in the radius and mass estimates from the \dnu\ scaling errors. The same exercise but with \dnu\ values calculated from mode frequencies will be performed later in the paper. However, for now we just want to examine the effects of the errors due to \numax\ scaling deviations. The results are shown in Figs.~\ref{fig:raderr}(a) and \ref{fig:masserr}(a). As can be seen, there are
systematic errors in both mass and radius estimates. Errors in both estimates are
reduced substantially when Eqs.~\ref{eq:radius}\ and \ref{eq:mass}\ are modified to
include the effect of the mean molecular weight and $\Gamma_1$, i.e.,
\begin{equation}
\frac{R}{R_\sun} = \left(\frac{\nu_{\rm max}}{\nu_{{\rm max},\sun}}\right) 
\left(\frac{\Delta \nu}{\Delta \nu_{\sun}}\right)^{-2} 
\left(\frac{T_{\mathrm{eff}}}{T_{\mathrm{eff,\sun}}}\right)^{1/2}
\left(\frac{\mu}{\mu_\odot}\right)^{-1/2} \left(\frac{\Gamma_1}{\Gamma_{1,\odot}}\right)^{-1/2},
\label{eq:radius2}
\end{equation}
and
\begin{equation}
\frac{M}{M_\sun} = \left(\frac{\nu_{\rm max}}{\nu_{{\rm max},\sun}}\right)^{3} 
\left(\frac{\Delta \nu}{\Delta \nu_{\sun}}\right)^{-4} 
\left(\frac{T_{\rm eff}}{T_{{\rm eff},\sun}}\right)^{3/2}
\left(\frac{\mu}{\mu_\odot}\right)^{-3/2} \left(\frac{\Gamma_1}{\Gamma_{1,\odot}}\right)^{-3/2}.
\label{eq:mass2}
\end{equation}
The errors in results obtained with these expressions are shown in 
Figs.~\ref{fig:raderr}(b) and \ref{fig:masserr}(b).
\begin{figure}
\epsscale{0.95}
\plotone{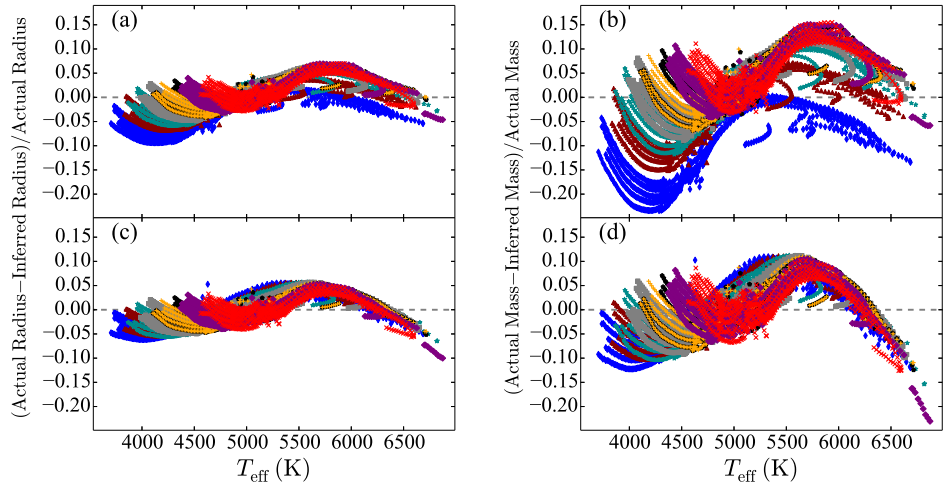}
\caption{The combined effect of the deviation of both \dnu\ and \numax\ on radius (a,c), and mass (b,d) estimates. The upper panels (a,b) show the fractional differences using the original scaling relation and the lower panels (c,d) show the deviations once the $\mu$ and $\Gamma_1$ terms are taken into account. Symbols and colors correspond to metallicities as indicated in Fig.~\ref{fig:s}.}
\label{fig:raderr2}
\end{figure}

The total error in radius and mass estimates obtained using the scaling laws are of course a combination of errors in the \dnu\ scaling relation as well as the \numax\ scaling relation. To determine what that is, instead of determining radius and mass using input \dnu\ values calculated using the scaling relation, we calculated the input \dnu\ using the
frequencies of $\ell=0$ modes assuming Gaussian weights around \numax\ with FWHM of $0.66 \nu_{\mathrm{max}}^{0.88}$ as from \cite{Mosser2012}. Once the value of $\Delta \nu$ from mode frequencies was calculated for each model, Eq.~\ref{eq:radius2}\ and \ref{eq:mass2} were again used to determine the error in radius and mass estimates. The results are show in Fig.~\ref{fig:raderr2} which also shows the errors in radius and mass estimates when the $\mu$ and $\Gamma_1$ terms are not included. Including the $\mu$ and $\Gamma_1$ terms helps reduce the deviations somewhat, but there is still substantial error in mass ($\pm 10-15$\%) and radius ($\pm 5$\%).

\newpage
\section{Discussion and conclusions}
\label{sec:concl}
We used a large set of models to test how well the \numax\ scaling holds,
and find that just as in the case of \dnu, there are significant departures
from the scaling law. The largest source of the deviation is the neglect
of the mean molecular weight and $\Gamma_1$ terms when approximating the acoustic cut-off frequency. The deviations in the scaling relations cause systematic errors in estimates
of \logg, mass, and radius. The errors in \logg\ are however, well within
errors caused by data uncertainties and are therefore not a big cause for
concern, except at extreme metallicities.

The results from our work would suggest we should start using the
$\mu$ and $\Gamma_1$ terms explicitly in the scaling relation, as in Eq. \ref{eq:numaxgamma}. Additionally, when using the scaling relations to determine radius and mass the $\mu$ and $\Gamma_1$ terms should be included, as in Eqs. \ref{eq:radius2} and \ref{eq:mass2}. For stellar models, ideally the best method is to use the actual value of $\mu$ and $\Gamma_1$ calculated in each model. For models where the values can be determined the modified scaling relation can be easily implemented. For models where $\Gamma_1$ is not readily accessible, we suggest still including the $\mu$ term in the \numax\ the scaling relation, which as seen in Fig.~\ref{fig:mures} is an improvement over the traditional \numax\ scaling relation. 

For observational data, incorporating these terms is not as straight forward. Even ignoring the $\Gamma_1$ term and just determining $\mu$ for an observed star is complicated. One possible way to implement the $\mu$ term into the scaling relation for observed stars would be to create stellar models and estimate the value of $\mu$ in this manner. For observed stars implementing the modified \numax\ scaling relation into the direct method (Eqs.~\ref{eq:radius2} and \ref{eq:mass2}) is not recommended due to the difficulty of determining the $\Gamma_1$ and $\mu$ terms from observational data. However, for observed stars a grid based method gives more precise estimates of radius and mass and should be used over the direct method. So, the difficulty in applying this result to observed stars is less critical.

Furthermore,  we should treat the \numax\ scaling the way we have begun to treat \dnu\ scaling, i.e., either calculate corrections to the relation or determine a reference \numax\ that depends on \teff\ to replace $\nu_{{\rm max},\odot}$ as the constant of
proportionality. For the non-diffusion Eddington atmosphere models a correction formula as a function of $T_{\mathrm{eff}}$ is provided in Appendix \ref{sec:temp_correction}. For grid-based modeling, we would suggest that \numax\ for the grid of models be calculated from the ratio
$\nu_{\rm ac}/\nu_{{\rm ac},{\rm SSM}}$ to avoid most of the systematic errors.

\acknowledgments The authors would like to thank Joseph R. Schmitt for the
use of some of the software he had written and Andrea Miglio for helpful comments and suggestions.
This work has been supported by NSF grant AST-1514676 and NASA grant NNX16AI09G to SB. WJC, GRD, and YE acknowledge the support of the UK Science and Technology Facilities Council (STFC). Funding for the Stellar Astrophysics Centre is provided by The Danish National Research Foundation (Grant DNRF106). 
\software{YREC \citep{Demarque2008}}
\vfill

\appendix
\section{Fractional Difference Between the Original and Modified Scaling Relation as a Function of [F$\mathrm{\MakeLowercase{e}}$/H]}
\label{sec:Determine_mu}

\begin{figure*}
\epsscale{0.65}
\plotone{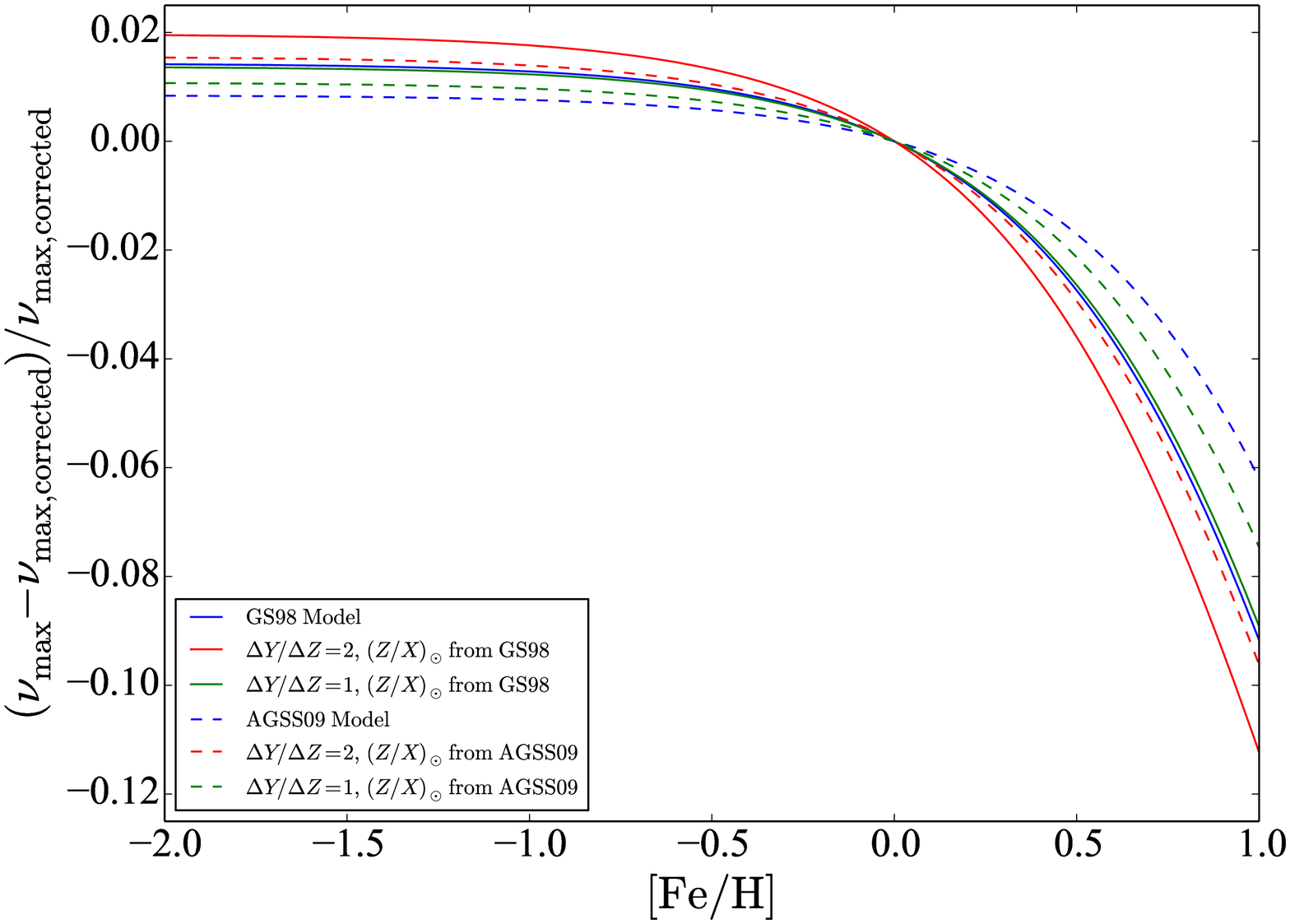}
\caption{The fractional difference between the traditional $\nu_{\mathrm{max}}$ value and the $\mu$ corrected $\nu_{\mathrm{max}}$ as a function of $\mathrm{[Fe/H]}$ for different values of $\Delta Y/\Delta Z$ and $(Z/X)_\sun$.
\label{fig:mu_correction_fig}
}
\end{figure*}

The effects of the $\mu$ correction can be seen if we plot the fractional difference between the traditional value of $\nu_{\mathrm{max}}$, as in Eq.~\ref{eq:numax}, and the value of $\nu_{\mathrm{max,corrected}}$ which includes the $\mu$ term as in Eq.~\ref{eq:numaxmu}. So, examining $\frac{\nu_{\mathrm{max}} - \nu_{\mathrm{max,corrected}} }{\nu_{\mathrm{max,corrected}}}$. By comparing Eq.~\ref{eq:numax} and Eq.~\ref{eq:numaxmu} it can be seen that 
\begin{equation}
\frac{\nu_{\mathrm{max}} - \nu_{\mathrm{max,corrected}}}{\nu_{\mathrm{max,corrected}}}=(\mu/\mu_{\sun})^{-1/2}-1.
\label{eq:mu_correction}
\end{equation}
The fractional difference between the traditional  $\nu_{\mathrm{max}}$ and $\nu_{\mathrm{max,corrected}}$ is shown in Fig.~\ref{fig:mu_correction_fig} as a function of $\mathrm{[Fe/H]}$ for different values of $\Delta Y/\Delta Z$ and $(Z/X)_\sun$.

\section{$\nu_{\mathrm{\MakeLowercase{max}}}$ Correction as a Function of Temperature}
\label{sec:temp_correction}

\begin{figure*}
\epsscale{0.65}
\plotone{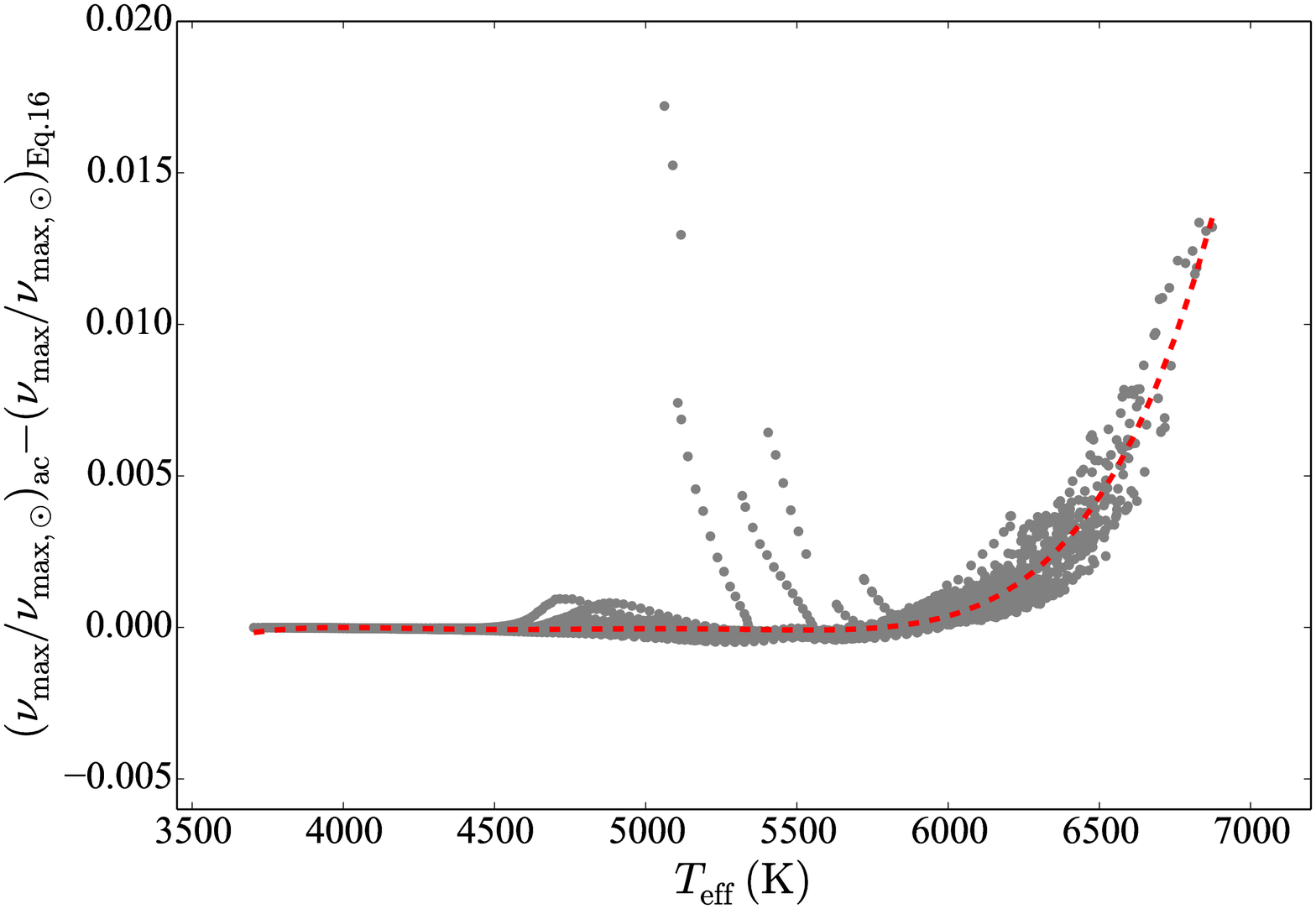}
\caption{The difference between $\nu_{\mathrm{max}}$ calculated from the acoustic-cutoff frequency and $\nu_{\mathrm{max}}$ calculated using Eq. \ref{eq:numaxgamma} including the $\mu$ and $\Gamma_1$ terms. The gray points show the non-diffusion Eddington atmosphere models and the red dashed line is a polynomial line of best fit (see Eq. \ref{eq:Temp_correction}).}
\label{fig:Temp_correction_fig}
\end{figure*}

Here we provide a correction formula for the non-diffusion Eddington models from Fig. \ref{fig:s}, solely as a function of $T_\mathrm{eff}$. For the Eddington atmosphere models, Figure \ref{fig:Temp_correction_fig} plots the difference between $\nu_{\mathrm{max}}$ determined from the acoustic-cutoff frequency and $\nu_{\mathrm{max}}$ determined using Eq.~\ref{eq:numaxgamma} (which includes the $\mu$  and $\Gamma_1$ terms) as a function of $T_\mathrm{eff}$. A fifth order polynomial was fit to the data, giving the relationship,
\begin{multline}
\left(\frac{\nu_{\mathrm{max}}}{\nu_{\mathrm{max,\sun}}}\right)_{\mathrm{ac}}-\left(\frac{\nu_{\mathrm{max}}}{\nu_{\mathrm{max,\sun}}}\right)_{\mathrm{Eq. 16}}= (4.5033 \times 10^{-19}) T_{\mathrm{eff}}^5 -(1.0781 \times 10^{-14}) T_{\mathrm{eff}}^4 + (1.0275 \times 10^{-10}) T_{\mathrm{eff}}^3 \\ -(4.8727 \times 10^{-7}) T_{\mathrm{eff}}^2 + (0.0011496) T_{\mathrm{eff}} -(1.0794).
\label{eq:Temp_correction}
\end{multline}
While this correction formula may be useful, the best and most accurate method to apply the $\mu$ and $\Gamma_1$ corrections is to calculate $\mu$ and $\Gamma_1$ for individual models and apply Eq.~\ref{eq:numaxgamma} to calculate $\nu_{\mathrm{max}}$. 

\bibliography{testing_nu_max}

\end{document}